\documentclass[tbtags,reqno]{amsart}
\numberwithin{equation}{section}

\newcommand{\abs}[1]{\lvert#1\rvert}

\begin{document}

\title[Overcomplete free energy functional for $D=1$ particle systems]
{Overcomplete free energy functional for $D=1$ particle systems with next
neighbor interactions}

\author{Christian Tutschka}
\address{Departamento de Matem\'aticas, Universidad Carlos III de Madrid,
Avenida de la Universidad 30, 28911 Legan\'es, Spain}
\thanks{C.T.\ thanks Karl Grill for helpful conversations. He is grateful to
the Austrian Science Foundation for financial support (Erwin Schr\"odinger
Grant \#2076)}

\author{Jos\'e A.\ Cuesta}
\address{Departamento de Matem\'aticas, Universidad Carlos III de Madrid,
Avenida de la Universidad 30, 28911 Legan\'es, Spain}
\thanks{J.A.C.\ acknowledges the support of the Direcci\'on General de
Investigaci\'on (Spain) through project BFM2000-0004}

\begin{abstract}
We deduce an overcomplete free energy functional for $D=1$ particle systems
with next neighbor interactions, where the set of redundant variables are the
local block densities $\varrho_i$ of $i$ interacting particles. The idea is to
analyze the decomposition of a given pure system into blocks of $i$
interacting particles by means of a mapping onto a hard rod mixture. This
mapping uses the local activity of component $i$ of the mixture to control
the local association of $i$ particles of the pure system. Thus it identifies
the local particle density of component $i$ of the mixture with the local
block density $\varrho_i$ of the given system. Consequently, our overcomplete
free energy functional takes on the hard rod mixture form with the set of
block densities $\varrho_i$ representing the sequence of partition functions
of the local aggregates of particle numbers $i$. The system of equations for
the local particle density $\varrho$ of the original system is closed via a
subsidiary condition for the block densities in terms of $\varrho$. Analoguous
to the uniform isothermal-isobaric technique, all our results are expressible
in terms of effective pressures. We illustrate the theory with two standard
examples, the adhesive interaction and the square-well potential. For the
uniform case, our proof of such an overcomplete format is based on the
exponential boundedness of the number of partitions of a positive integer
(Hardy-Ramanujan formula) and on Varadhan's theorem on the asymptotics of a
class of integrals. We also discuss the applicability of our strategy in
higher dimensional space, as well as models suggested thereof.
\end{abstract}

\maketitle

\section{Introduction}

The era of the class of exactly solvable $D=1$ particle systems with next
neighbor interactions in classical density functional theory (for an overview
see, e.g., \cite{p96}) began in 1976, when Percus derived the free energy
density functional for hard rods in an arbitrary external potential field
\cite{p76}. Soon afterwards, he presented the inverse operator format
\cite{p82}. This formalism not only allowed him to simplify the derivation of
the hard rod functional, but also led him to the solution of the sticky core
model, in which the interaction has a $0$-range attractive component.
Moreover, he showed that---analoguous to the uniform case---the adhesive
interaction is the only potential with an attractive component, where an
explicit solution is possible. The generalization of the pure hard core result
to an additive mixture of hard rods was established by Vanderlick, Davis, and
Percus in 1989 \cite{vdp89}. Already a few years later, Brannock and Percus
constructed along Wertheim's theory of local association (for a stringent
review of Wertheim's approach see \cite{p96}) a free energy density functional
for multicomponent systems with arbitrary next neighbor interactions
\cite{bp96}. One of the specific examples treated was the one-component case
of the prototypical square-well interaction, called Herzfeld Goeppert-Mayer
(HGM) system \cite{hg34}. In abstracting from this Wertheim local association
free energy density functional format, Percus invented in 1997 the
overcompleteness technique \cite{p97}. In this paper, we present a particular
realization of this strategy.

The essential idea of an overcomplete description \cite{p97} of thermodynamic
functionals is to introduce a set of additional variables (over the basic
local particle density $\varrho$) such that with respect to a direct
evaluation the overcomplete format has a simpler formal structure and is
physically more transparent. Thus, apart from the purely mathematical
problem of deriving such a representation from first-principles, the second
difficulty \cite{p97} is to give a physical interpretation for the redundant
variables.

In our approach we use an observation of Kierlik and Rosinberg \cite{kr92} to
overcome the latter difficulty. Their idea, although stated originally only for
nonuniform dimerizing hard rods (for the uniform case see, e.g., \cite{hg34}),
consists in interpreting a $D=1$ particle system with next neighbor
interactions as composed of blocks (clusters \cite{kt00}, molecules
\cite{hg34,kr92}, or superparticles \cite{bp96}) of $i$ interacting particles
(monomers, dimers, trimers,~\dots) and to use the corresponding local block
densities $\varrho_i$ as the set of additional variables. Here two particles
are understood to interact, if in the standard decomposition of the Boltzmann
factor into a reference hard core and an attractive part, \eqref{sd}, the
interaction between the two particles is described by the latter one. Since
the blocks of local aggregates interact by construction via hard core
potentials, the free energy functional takes on the simple hard rod mixture
form. In such an overcomplete description the physical significance of the
redundant variables is given at the outset.

So our program is to analyze a given pure system with arbitrary next neighbor
interactions by means of a mapping onto a hard rod mixture. This mapping uses
the local activity of component $i$ of the mixture to represent the local
aggregates of $i$ particles of the pure system and so it identifies the local
particle density of component $i$ of the mixture with the local block density
$\varrho_i$ of the original system. Consequently, with the free energy of
hard rod mixture form, the set of local block densities represents the
sequence of partition functions of the local aggregates with particle numbers
$i$. We close the system of equations for the local particle density via a
relation between $\varrho$ and the block densities inferred through a trivial
identity of functional analysis.

Thus already implicitly formulated by Kierlik and Rosinberg \cite{kr92},
they settled this general program only for the special case of dimerizing
hard rods. They argued that, although this technique can be extended to the
more general case of sticky cores, where blocks of any number of particles
coexist, the possible difficulty of this method lies in the definition of
blocks for arbitrary next neighbor interactions. So they proceeded with
Wertheim's theory instead. They (i) showed that Wertheim's first-order
thermodynamic perturbation theory (TPT1) is exact for dimerizing hard rods as
well as for sticky cores and (ii) derived density functionals for these
systems with the simple hard rod mixture form through TPT1. Later on, the
exactness of the TPT1 also for sticky core mixtures was established by
Brannock and Percus \cite{bp96}.

Actually, already Brannock and Percus \cite{bp96} used the block
interpretation of Kierlik and Rosinberg \cite{kr92} to construct an
overcomplete free energy functional. But only formally, as a purely technical
intermediate step in their proof, because their aim was to infer density
functionals that take on the Wertheim local association form. However, in
such an overcomplete description the remarkable hard rod mixture form of the
free energy is obscured. As already stated, again Kierlik and Rosinberg
\cite{kr92} first revealed that at least for dimerizing hard rods and for
sticky cores, functionals of the simple hard rod mixture form underly the
Wertheim functionals.

Similarly, also with respect to the relative density format of Percus
\cite{p97}, another earlier and formally alike equivalent overcomplete
description, where the redundant variables are partition function densities,
our realization has a simpler formal characterization (in terms of a
functional of hard rod mixture form) and is physically more transparent (with
the block densities as the basic elements of the theory). Moreover, all our
main results possess higher dimensional analogues, so that our overcomplete
description immediately suggests $D>1$ models as well.

Formally, we start from the (configurational) isothermal measure (or the Gibbs
measure thereof) for a particle system on the line $\mathbb R$ with next
neighbor interactions, given by the density
\begin{equation}\label{id}
        \prod_{k=1}^{n-1} e(q_{k+1},q_k)\prod_{k=1}^n\exp[-\beta U(q_k)]
\end{equation}
with respect to the usual Lebesgue measure. Here, $n$ denotes the number of
particles, their positions being $q_k\in\mathbb R$, $U$ is the external
potential, and the translation-invariant interaction $V$ is represented by
the Boltzmann factor (at reciprocal temperature $\beta$)
\begin{equation}\label{bf}
        e(y,x)=\begin{cases}
                \exp[-\beta V(y-x)]     &\text{for $y \geq x$},\\
                0                       &\text{for $y < x$}.
               \end{cases}
\end{equation}
We assume that $V$ has a hard core of diameter $a>0$. Hence if we restrict
ourselves to the next neighbor case, the range of $V$ is $2a$, i.e.,
$V(x)\equiv 0$, $x\geq 2a$. Furthermore, we introduce the standard
decomposition (for a related analysis based on a different decomposition see,
e.g., \cite{kt00})
\begin{equation}\label{sd}
        e(y,x)=h_a(y-x)+f(y,x)
\end{equation}
with $h_a$ the Heaviside function shifted to the right by $a$ and
$f(y,x)=h_a(y-x)[e(y,x)-1]$, known as the attractive part of the Boltzmann
factor.

First, we will treat for the sake of completeness the uniform case within our
overcomplete description. This will also allow us to introduce the concept
most clearly. Our account will be inductive. We are going to develop the
theory along two standard examples, the sticky core model and the HGM system.
Second, we will lift our overcomplete format to the general nonuniform case
and illustrate the results with the two examples already examined for the
uniform case. Finally, we will indicate how the idea extends to higher
dimensional space and suggest $D>1$ models thereof.

\section{Uniform case}\label{uc}

In this section we introduce the concept of overcomplete description for the
free energy of uniform systems with arbitrary next neighbor interactions. We
derive such a representation from first-principles as $n\to\infty$ based on
the exponential boundedness of the number of partitions of a positive integer
$n$ (Hardy-Ramanujan formula, see, e.g., \cite{a94}) and on Varadhan's theorem
on the asymptotic behavior of a class of measures (see, e.g., \cite{e85}).

Formally, we assume $U(x)\equiv 0$, $x\in\mathbb R$. Then the total isothermal
measure on $(0,l)^n$, under the condition that the particles be reflected
elastically at the endpoints of the interval $(0,l)$ (the volume
$\abs{(0,l)}=l$ is called the length of the system), becomes
\begin{equation}\label{fepp}
        \exp[-n\beta f_n(l)]=
        \int_{(0,l)^n}\prod_{k=1}^{n-1} e(q_{k+1},q_k)\,d(q_1,\dotsc,q_n),
\end{equation}
where we have introduced $f_n$ known as the (configurational) $n$-particle
free energy per particle.              

For the sake of explicitness and to make contact with a previous paper
\cite{kr92}, we start our presentation with the sticky core model.

\subsection{Adhesive interaction}

Adhesive particle systems are systems composed of hard rods with an additional
attractive interaction on the boundary of the particles, which allows for
local association. Thus an adhesive system may be represented as a limiting
case of a HGM system. In such systems, the interaction is
\begin{equation}\label{i.hgm}
        V(x)=\begin{cases}
                +\infty&\text{for $x\in[0,a)$,}\\
                -{\mathcal E}\chi_{[0,1)}\big(\frac{x-a}{d}\big)
                &\text{for $x\in[a,\infty)$,}
                \end{cases}
\end{equation}
where $d$ is the width of the well (note that we have restricted ourselves to
$d\leq a$), $\mathcal E$ the interaction strength, and $\chi_A$ denotes the
characteristic function of the set $A$. Then---on inserting \eqref{i.hgm} into
\eqref{sd} via \eqref{bf}---adhesive systems are characterized by the limit
\begin{equation}\label{ap.ai}
        \lim_{\begin{subarray}{l}\mathcal E\to\infty\\d\to 0\end{subarray}}
        f(y,x)|_{d e^{\beta\mathcal E}\to\lambda}=\lambda\,\delta_a(y-x)
\end{equation}
with $\delta_a=h_a'$, $\lambda=\gamma a$, and $\gamma\in\mathbb R_+$ being the
stickiness parameter of the interaction.                    

Hence, due to the $0$-range attractive component \eqref{ap.ai}, our
overcomplete description for the $n$-particle sticky core model translates to
the decomposition into blocks (monomers, dimers, trimers,~\dots,~$n$-mers) of
pure hard core particles with diameters $(a_1,\dotsc,a_n)$, $a_i=ia$, and to
the introduction of the $n$ block densities as the set of redundant variables.
Therefore $f_n$ is defined on an enlarged space with $n$ additional dimensions.
Consequently, the free energy is calculated by means of a minimum principle,
\eqref{u.od.ai.fe}.

Formally, upon substituting \eqref{ap.ai} into \eqref{fepp} via \eqref{sd},
$f_n$ can be interpreted as a sum over a binary tree with $2^{n-1}$ branches.
Consider the contribution of an arbitrary branch, i.e., an arbitrary
convolution sequence of $h_a$ and $\lambda\,\delta_a$, to this sum. For
$y,x\in(0,l)$ denote by a block (or cluster \cite{kt00}, molecule
\cite{hg34,kr92}, or superparticle \cite{bp96}) $b_i$ of $i\in\{1,\dots,n\}$
interacting particles the convolution sequence
\begin{equation}\label{u.bi.ai}
        b_i(y,x)=\begin{cases}
                h_a(y-x)                          &\text{for $i=1$},\\
                b_{i-1}\ast\lambda\,\delta_a(y-x) &\text{for $1<i\leq n$}
                 \end{cases}
\end{equation} 
with $a\ast b(y,x)=\int_0^l a(y-z)b(z-x)\,dz$. Let $k_i$ be the number of
times $b_i$ (modulo $h_a$ at the right boundary) appears in such a branch. For
each $n\geq 2$ define the set
\begin{equation}
        A_n=\bigg\{k=(k_1,\dotsc,k_n):\,k_i\in\{0,\dotsc,n\}\wedge
        \sum_{i=1}^n i k_i=n\bigg\}.
\end{equation}
Fix $k\in A_n$. Then, on using the commutativity of the convolution (for
another application of this elementary property see, e.g., \cite{kt00}), we
conclude that each of the
\begin{equation}
        C(n,k)=\dfrac{\left(\sum_{i=1}^n k_i\right)!}{k_1!\dotsm k_n!}
\end{equation}
branches characterized by the vector $k$ contributes the same value to $f_n$.
Again through the commutativity of the convolution, we can rearrange the blocks
according to their length, starting with the group of $k_1$ particles of
length $a$ next to the origin, so that the total isothermal measure reduces to
\begin{equation}
        \exp[-n\beta f_n(l)]=
        \sum_{k\in A_n}C(n,k)\,\lambda^{\sum_{i=1}^n(i-1)k_i}
        \exp[-n\beta f_{n,k}(l)]
\end{equation}
with
\begin{align}
        \mathcal{L}\{\exp[-n\beta f_{n,k}(l)]\}(s)&=
        \dfrac{1}{s}\exp(as)\,s
        \prod_{i=1}^n\left[\dfrac{\exp(-ias)}{s}\right]^{k_i}
        \dfrac{1}{s}\nonumber\\
        &=\dfrac{1}{s}
        \dfrac{\exp\left[-\left(\sum_{i=1}^n ik_i-1\right)as\right]}
        {s^{\sum_{i=1}^n k_i-1}}
        \dfrac{1}{s},
        \quad k\in A_n,
\end{align}
where $\mathcal{L}\{\exp[-n\beta f_{n,k}(l)]\}(s)$ denotes the Laplace
transform of $\exp[-n\beta f_{n,k}(l)]$ with respect to $l$ and we used the
relative positions
\begin{equation}\label{rp}
        x_0=q_1,\quad x_k=q_{k+1}-q_k\text{ for }1\leq k<n,\quad x_n=l-q_n
\end{equation}
as well as \eqref{u.bi.ai}. On transforming back, we infer
\begin{equation}\label{u.od.ai}
        \exp[-n\beta f_n(l)]=
        h_{(n-1)a}(l)\sum_{k\in A_n}C(n,k)\,\lambda^{\sum_{i=1}^n(i-1)k_i}
        \,\dfrac{\left[l-\left(\sum_{i=1}^n ik_i-1\right)a\right]
        ^{\sum_{i=1}^n k_i}}{\left(\sum_{i=1}^n k_i\right)!}.
\end{equation}

Finally, let us analyze the leading order asymptotic behavior of $f_n$ for
the nontrivial case $l>(n-1)a$. Denote by $\pi(n)$ the number of partitions
of a positive integer $n$. Since $\abs{A_n}=\pi(n)$, it follows by the leading
term of the asymptotic series of the Hardy-Ramanujan formula (see, e.g.,
\cite{a94}) that as $n\to\infty$
\begin{equation}\label{hrf}
        \abs{A_n}\asymp\dfrac{1}{2\pi\sqrt{2}}\,\dfrac{d}{dn}
        \left[\dfrac{\exp\bigg(
        \dfrac{2\pi}{\sqrt{6}}\sqrt{n-\dfrac{1}{24}}\bigg)}
        {\sqrt{n-\dfrac{1}{24}}}\right].
\end{equation}
Hence the number of terms in the sum \eqref{u.od.ai} is exponentially bounded. 
Thus one may determine the asymptotic behavior of $f_n$ by the asymptotic
behavior of the largest summand in \eqref{u.od.ai}. This is a standard
argument of large deviation theory (see, e.g., \cite{e85}). Therefore, if we
use a weak form of Stirling's formula, $\ln(n!)=n\ln n -n +O(\ln n)$, we find
as $n\to\infty$ 
\begin{equation}\label{lemma}
        \dfrac{1}{n}\ln C(n,k)=\sum_{i=1}^n\dfrac{k_i}{n}
        \ln\bigg(\sum_{j=1}^n\dfrac{k_j}{n}\bigg)
        -\sum_{i=1}^n\dfrac{k_i}{n}\ln\dfrac{k_i}{n}
        +O\left(\dfrac{\ln n}{n}\right),
\end{equation}
and so, on introducing the particle density $\rho^{-1}=l/n$ with
$\rho<a^{-1}$ as $n\to\infty$ by $l>(n-1)a$, via \eqref{hrf} that
\begin{equation}\label{u.od.ai.fe.fa}
        \lim_{n\to\infty}\beta f_n(\rho)
        =\lim_{n\to\infty}\min_{k\in A_n}
        \bigg\{\sum_{i=1}^n\dfrac{k_i}{n}\bigg[\ln\dfrac{k_i}{n}-1
        -\ln\lambda^{i-1}
        -\ln\bigg(\rho^{-1}
        -a\sum_{i=1}^n i\dfrac{k_i}{n}\bigg)
        \bigg]\bigg\}.
\end{equation}
By the properties of the sets $\{A_n\}_{n\geq 2}$, of the set
$A=\big\{t=(t_1,\dotsc):\,t_i\geq 0 \wedge\sum_{i=1}^{\infty} i t_i=1\big\}$,
and of the function on the right hand side (RHS) of \eqref{u.od.ai.fe.fa}, we
finally infer
\begin{equation}\label{u.od.ai.fe}
        \lim_{n\to\infty}\beta f_n(\rho)\equiv\beta f_{\infty}(\rho)
        =\min_{c\in A}\bigg\{\sum_{i=1}^{\infty} c_i\left[
        \ln\dfrac{c_i}{\lambda^{i-1}}-1
        -\ln\left(\rho^{-1}-a\right)\right]\bigg\},
\end{equation}
where $f_{\infty}$ is the free energy per particle. This proves (4.2)--(4.4)
of \cite{kr92}.

Fix $\gamma,a\in\mathbb R_+$, $\rho\in\left(0,a^{-1}\right)$. The function
appearing on the RHS of \eqref{u.od.ai.fe} is convex. Hence it attains its
minimum at the unique value $c=(c_1,\dotsc)$ with
\begin{equation}\label{u.od.ai.c}
        c_i=\begin{cases}
                \dfrac{1}
                {\left(1+\lambda\theta\right)^2}        &\text{for $i=1$},\\
                \cfrac{c_1}{\bigg(1+\cfrac{1}
                {\lambda\theta}\bigg)^{i-1}}            &\text{for $i>1$}.
             \end{cases}
\end{equation}
Here, while it is possible to write $c$ directly as a function of $\rho$, we
preferred to introduce for later reference the pressure $p$ via the
thermodynamic relation
\begin{equation}\label{tr}
        \beta p(\rho)=
        \rho^2\dfrac{\partial}{\partial\rho}\beta f_{\infty}(\rho)
\end{equation}
with $\theta\equiv\beta p(\rho)$. Thus we have by
\begin{gather}\label{u.od.ai.mc}
        \sum_{i=1}^{\infty}c_i=\dfrac{1}{1+\lambda\theta}
\intertext{that}\label{u.eos.ai.i}
        \theta(1+\lambda\theta)=\dfrac{1}{\rho^{-1}-a},\\
\intertext{or explicitly}\label{u.eos.ai}
        \theta=\dfrac{1}{2\lambda}
        \left(-1+\sqrt{1+\dfrac{4\lambda}{\rho^{-1}-a}}\right).\\
\intertext{Therefore $f_{\infty}$ simplifies to}\label{u.ai.fe}
        \beta f_{\infty}(\rho)=-\ln\left(1+\lambda\theta\right)
        -\theta\left(\rho^{-1}-a\right)+\ln\theta,
\end{gather}
which is the known result. Formula \eqref{u.od.ai.c} is equivalent to the
findings (4.5) and (4.7) of \cite{kr92}.

As will be shown in Section \ref{nc}, the remarkable hard rod mixture form of
$f_{\infty}$ as given by \eqref{u.od.ai.fe}, is an instance of a general
result that even holds true in the context of nonuniformity.

\subsection{Arbitrary next neighbor interactions}

So far, we treated particle systems with adhesive interactions. Essentially,
the full and explicit analysis was possible because of the contact nature of
the adhesive potential. When arbitrary next neighbor interactions are present,
only the minimum of the length of a block $b_i$ of $i$ interacting particles
is $ia$. The purpose of the present subsection is to establish our
overcompleteness technique for the general uniform case.

Formally, our proof of such an overcomplete description for uniform systems
with arbitrary next neighbor interactions as $n\to\infty$ is based on the
Hardy-Ramanujan formula, \eqref{hrf}, and on an application of Varadhan's
theorem on the asymptotic behavior of a class of measures (see, e.g.,
\cite{e85}). 

Consider first the prototypical HGM system. Then, upon substituting
\eqref{i.hgm} into \eqref{sd} via \eqref{bf}, the total isothermal measure on
$(0,l)^n$ is completely characterized by
\begin{equation}\label{ap.hgm}
        f(y,x)=\lambda\,h_{a,d}(y-x)
\end{equation}
with $h_{a,d}=h_a-h_{a+d}$ and $\lambda=e^{\beta\mathcal E}-1$. Hence, as in
the deduction of our overcomplete description for the adhesive interaction, we
start the proof by analyzing an arbitrary branch of the binary tree associated
with $f_n$. Let us again decompose such a branch into $k_i$ (modulo $h_a$ at
the right boundary) blocks $b_i$ of $i\in\{1,\dots,n\}$ interacting particles,
now defined by the convolution sequence [cf.\ \eqref{u.bi.ai}]
\begin{equation}\label{u.bi.hgm}
        b_i(y,x)=\begin{cases}
                h_a(y-x)                         &\text{for $i=1$},\\
                b_{i-1}\ast\lambda\,h_{a,d}(y-x) &\text{for $1<i\leq n$}.
                 \end{cases}
\end{equation}
Then, via \eqref{rp} as well as \eqref{u.bi.hgm} and on using the
commutativity of the convolution, we conclude that
\begin{equation}\label{u.od.hgm}
        \exp[-n\beta f_n(l)]=
        \sum_{k\in A_n}C(n,k)\,\lambda^{\sum_{i=1}^n(i-1)k_i}
        \exp[-n\beta f_{n,k}(l)]
\end{equation}
with
\begin{align}
        \mathcal{L}\{\exp[-n\beta f_{n,k}(l)]\}(s)&=
        \dfrac{1}{s}\exp(as)\,s
        \prod_{i=1}^n\left\{\dfrac{\exp(-ias)}{s}
        \left[\dfrac{1-\exp(-ds)}{s}\right]^{i-1}\right\}^{k_i}
        \dfrac{1}{s}\nonumber\\
        &=\dfrac{1}{s}\left[\dfrac{\exp(-as)}{s}\right]^{n-1}
        \left[1-\exp(-ds)\right]^{\sum_{i=1}^n(i-1)k_i}
        \dfrac{1}{s},
        \quad k\in A_n.
\end{align}
       
Next, let as usual $l=n\rho^{-1}$, and consider the asymptotic behavior of
$f_n$ for the nontrivial case $\rho^{-1}>a$. Since $\abs{A_n}$ is exponentially
bounded as $n\to\infty$ by the Hardy-Ramanujan formula, \eqref{hrf}, the
asymptotics of $f_n$ is governed by the asymptotic behavior of the largest
summand in \eqref{u.od.hgm}. Thus, by the formal structure of \eqref{u.od.hgm},
the analysis of $f_n$ as $n\to\infty$ reduces to an asymptotic evaluation of
$f_{n,k}$. If we combine such an evaluation of $f_{n,k}$ as $n\to\infty$ by
means of Varadhan's theorem (see, e.g., \cite{e85}) with \eqref{lemma}, we
finally infer for $\beta\in(0,\infty)$
\begin{align}\label{u.od.hgm.fe}
        \beta f_{\infty}(\rho)
        &=\min_{c\in A}\bigg(\sum_{i=1}^{\infty}c_i
        \ln\dfrac{c_i}{\lambda^{i-1}\sum_{j=1}^{\infty}c_j}
        -\min_{\theta\in\mathbb R_+}
        \bigg\{\theta\left(\rho^{-1}-a\right)\nonumber\\
        &-\ln\theta
        +\bigg(1-\sum_{i=1}^{\infty}c_i\bigg)
        \ln\left[1-\exp(-\theta d)\right]
        \bigg\}\bigg).
\end{align}
The case $\beta=\infty$ has to be treated separately (see, e.g., \cite{kt00}).

Fix $a,\rho^{-1},\mathcal{E},d,\beta\in\mathbb R_+$ such that $d\leq a$ and
$\rho^{-1}>a$. Then the function appearing on the RHS of \eqref{u.od.hgm.fe}
is convex. Hence the minimum is attained at the unique value
$c=(c_1,\dotsc)$ with
\begin{equation}\label{u.od.hgm.c}
        c_i=\begin{cases}
                \dfrac{1}
                {\left\{1+\lambda\left[1-\exp(-\theta d)\right]\right\}^2}
                                                        &\text{for $i=1$},\\
                \cfrac{c_1}{\bigg\{1+\cfrac{1}
                {\lambda\left[1-\exp(-\theta d)\right]}\bigg\}^{i-1}}
                                                        &\text{for $i>1$},
             \end{cases}
\end{equation}
and thus
\begin{gather}\label{u.od.hgm.mc}
        \sum_{i=1}^{\infty}c_i=
        \dfrac{1}{1+\lambda\left[1-\exp(-\theta d)\right]}.\\
\intertext{Therefore $f_{\infty}$ simplifies for $\beta\in(0,\infty)$ to}
\label{u.hgm.fe}
        \beta f_{\infty}(\rho)
        =-\ln\left\{1+\lambda\left[1-\exp(-\theta d)\right]\right\}
        -\theta\left(\rho^{-1}-a\right)+\ln\theta,
\end{gather}
which is the known result. In the sticky limit
$\mathcal E\to\infty\wedge d\to0$ such that
$d e^{\beta\mathcal E}\to\gamma a$, expressions \eqref{u.hgm.fe} respective
\eqref{u.od.hgm.c} reduce to \eqref{u.ai.fe} respective \eqref{u.od.ai.c}. An
approximation of \eqref{u.od.hgm.c} was given by Herzfeld and Goeppert-Mayer
already in 1934 \cite{hg34}.

Moreover, through \eqref{tr} and \eqref{u.od.hgm.fe}, we have the
interpretation
\begin{equation}
        \theta\equiv\beta p(\rho),
\end{equation}
and so, upon using \eqref{u.od.hgm.c} in \eqref{u.od.hgm.fe}, that the
pressure for $\beta\in(0,\infty)$ is uniquely determined by
\begin{equation}\label{u.eos.hgm.i}
        \rho^{-1}-a=\dfrac{1}{\beta p}
        -\cfrac{d}{\exp(\beta p d)\bigg(1+\cfrac{1}{\lambda}\bigg)-1}.
\end{equation}
Hence, $p$ is a differentiable function with range $\mathbb R_+$. In
particular, we have $\tfrac{\partial p}{\partial\rho}>0$.

Finally, since Varadhan's theorem allows us to analyze the asymptotic behavior
of a large class of total measures $f_{n,k}$, our scheme of proof for the HGM
system extends directly to (in this sense) arbitrary next neighbor
interactions. This completes our presentation of the uniform case.

\section{Nonuniform case}\label{nc}

We now generalize the ideas of Section \ref{uc} to the nonuniform case. Our
approach is formal. We assume that for a sufficiently large class of pair
interactions $V$ and external potentials $U$ the corresponding free energy
density functional exists and is convex. In fact, Chayes and Chayes
\cite{cc84} showed, that for our domain of interest of $D=1$ particle systems
with stable (see, e.g., \cite{r69}) hard core next neighbor interactions, the
set of external potentials for which a convex free energy density functional
exists is nonempty. Thus, our calculus solves indeed a well-posed inverse
problem for (in this sense) arbitrary next neighbor interactions.

We construct our overcomplete free energy functional through a reduction of
the original  problem to a readily solvable multicomponent inverse problem.
The blocks of $i$ interacting particles [cf.\ for the uniform case the
instances \eqref{u.bi.ai} respective \eqref{u.bi.hgm}] are controlled by an
effective local activity of component $i$ of the mixture. Hence, since the
blocks interact by definition via hard core potentials, the interaction
between components $i$ and $j$ of the mixture is represented by an effective
pure hard core Boltzmann factor. Thus, we reduce the inverse problem for
arbitrary next neighbor interactions to a type of polydisperse hard rod
mixture problem. Consequently, we first extend the technique of proof of
Vanderlick, Davis, and Percus \cite{vdp89} for an additive mixture of pure hard
cores in an external field to our case of polydispersity. Results for the
local block densities are given next. Finally, the isomorphism between the
original and our overcomplete description is established via a trivial
identity of functional analysis.

The presentation of the nonuniform case closes with the two examples already
treated at the uniform level. The especially suggestive formal structure of
these results provides evidence for our conjecture on a direct nonuniform
isothermal-isobaric representation.

\subsection{Overcomplete free energy functional format}

We start with the standard description of nonuniform systems in direct form.
So consider first the Gibbs measure (see, e.g., \cite{r69}) of \eqref{id} at
chemical potential $\mu$ (implicitly including the momentum contributions).
Denote the local activity by $z(x)=\exp\{\beta[\mu-U(x)]\}$. Then the total
Gibbs measure becomes \cite{p82} \big[in Dirac notation
$ab\equiv ab(y,x)=\int_{\mathbb{R}}a(y,z)b(z,x)\,dz\big]$
\begin{equation}\label{n.tgm.sd}
        \Xi=1+Iz(\mathbb{I}-ez)^{-1}I,
\end{equation}
where $\mathbb{I}$ is the identity, $I$ represents the constant $I(x)=1$, $z$
refers to the matrix with elements $z(y,x)=z(x)\delta(y-x)$, and $e$ is
given by \eqref{bf}. If $\lim_{x\to\pm\infty}z(x)=0$ sufficiently rapidly,
then $\Xi$ exists. Hence, on following Percus \cite{p82,p97}, the local
particle density $\varrho$ can be written as
\begin{gather}\label{n.pd}
        \varrho(x)=\dfrac{\delta\ln\Xi}{\delta\ln z(x)}
        =\dfrac{1}{\Xi}\Xi_i^+(x)z(x)\Xi^-(x)
\intertext{with}\label{n.pff.sd}
        \Xi^-=(\mathbb{I}-ez)^{-1}I,\quad\Xi^+=I(\mathbb{I}-ze)^{-1},
\end{gather}
and thus we have by \eqref{bf} as well as $\lim_{x\to\pm\infty}z(x)=0$ the
boundary conditions
\begin{subequations}\label{n.bc.sd}
\begin{align}
        \lim_{x\to-\infty}\Xi^-(x)&=\lim_{x\to+\infty}\Xi^+(x)=1,\\
        \lim_{x\to+\infty}\Xi^-(x)&=\lim_{x\to-\infty}\Xi^+(x)=\Xi,\\
        \lim_{x\to\pm\infty}\varrho(x)&=0.
\end{align}
\end{subequations}
The strategy is to determine $z$ as a functional of $\varrho$ for (in the above
sense) arbitrary $V$. The main stage to settle this inverse problem is the
concept of overcomplete description, formulated next.

Consider the Gibbs measure for a nonuniform multicomponent system on the line
$\mathbb R$ \cite{vdp89,p97}, where the local activity of component $i$ is
given by [henceforth a bar signifies a quantity within the overcomplete
description]
\begin{equation}\label{n.la.od}
        \bar{w}_i(x,l)=z(fz)^{i-1}(x+\tfrac{l}{2},x-\tfrac{l}{2}).
\end{equation}
Here, $x$ is the position of the block of $i$ interacting particles, $l+a$
its diameter with $l$ being the distance between the boundary particles, and
$f$ is brought in via \eqref{sd}. Hence definition \eqref{n.la.od}, as the
total isothermal measure of a $i$ particle system in a box
$(x-\tfrac{l}{2},x+\tfrac{l}{2})$ with fixed boundary particles and
Boltzmann factor $f$, controls the local aggregates. So the block description
[cf.\ for the uniform case the examples \eqref{u.bi.ai} respective
\eqref{u.bi.hgm}] is completed by the pure hard core Boltzmann factor between
components $i$ and $j$, given within the present parameterization by
\begin{equation}\label{n.bf.od}
        \bar{e}_{ij}(x,x',l,l')=h_{\tfrac{l+l'}{2}+a}(x-x').
\end{equation}
Then, on assuming that the indices of the blocks are determined statistically
\cite{vdp89,p97}, the total Gibbs measure for such a mixture can be written as
(cf.\ \cite{vdp89,p97})
\begin{equation}\label{n.tgm.od}
        \bar{\Xi}=1+\bar{I}^T\bar{w}(\mathbb{I}-\bar{e}\bar{w})^{-1}\bar{I},
\end{equation}
where $\bar{e}$ is the matrix with elements $\bar{e}_{ij}(x,x',l,l')$,
$\bar{w}$ represents $\bar{w}_{ij}(x,x',l,l')
=\bar{w}_i(x,l)\delta(x-x')\delta(l-l')\delta_{ij}$,
$\bar{I}_i(x)=1$ is the constant vector, and we used the Dirac-Einstein
notation $ab\equiv a_{ik}b_{kj}(y,x)
=\sum_{k=1}^{\infty}\int_{\mathbb{R}}a_{ik}(y,z)b_{kj}(z,x)\,dz$. A term by
term analysis of the sums \eqref{n.tgm.sd} and \eqref{n.tgm.od} shows
$\bar{\Xi}\equiv\Xi$. Hence we conclude
\begin{equation}\label{n.tgm.od.df}
        \Xi=1+\bar{I}^T\bar{w}(\mathbb{I}-\bar{e}\bar{w})^{-1}\bar{I},
\end{equation}
which is the desired overcomplete description in direct form. Similarly, upon
using $\sum_{i=1}^{\infty}(fz)^{i-1}=(\mathbb{I}-fz)^{-1}$, the equivalence of
\eqref{n.tgm.od} with the Brannock-Percus format \cite{bp96} can be formally
established. Finally, the local block densities are found to be
\begin{gather}\label{n.bd}
        \varrho_i(x,l)=\dfrac{\delta\ln\Xi}{\delta\ln\bar{w}_i(x,l)}
        =\dfrac{1}{\Xi}\bar{\Xi}^+(x,l)\bar{w}_i(x,l)\bar{\Xi}^-(x,l)
\intertext{with}\label{n.pff.od}
        \bar{\Xi}^-=(\mathbb{I}-\bar{e}\bar{w})^{-1}\bar{I},\quad
        \bar{\Xi}^+=\bar{I}^T(\mathbb{I}-\bar{w}\bar{e})^{-1}.
\end{gather}
Therefore by \eqref{n.bf.od} and again $\lim_{x\to\pm\infty}z(x)=0$ we have
for $l\in\mathbb R$
\begin{subequations}\label{n.bc.od}
\begin{align}
        \lim_{x\to-\infty}\bar{\Xi}^-(x,l)&=
        \lim_{x\to+\infty}\bar{\Xi}^+(x,l)=1,\\
        \lim_{x\to+\infty}\bar{\Xi}^-(x,l)&=
        \lim_{x\to-\infty}\bar{\Xi}^+(x,l)=\Xi,\\
        \lim_{x\to\pm\infty}\varrho_i(x,l)&=0.
\end{align}
\end{subequations}

Now, what makes the overcomplete description \eqref{n.tgm.od.df} more handy
than the original formulation \eqref{n.tgm.sd} is, that as for an additive
hard rod mixture $\bar{e}$ is of rank $1$ on index space (cf.\ \cite{p97}),
and hence can be immediately written as
\begin{equation}\label{n.rod}
        \bar{e}_{ij}=h_{\tfrac{l+a}{2}}\delta_{\tfrac{l'+a}{2}}
        =\delta_{\tfrac{l+a}{2}}h_{\tfrac{l'+a}{2}}.
\end{equation}
Correspondingly, the technique of proof of \cite{vdp89} for the free energy
functional of additive mixtures of hard rods goes indeed through, with only
minor modifications, for the case of our overcomplete free energy functional
for arbitrary next neighbor interactions.

The details are as follows. Introduce the one-dimensional auxiliary functions
(cf.\ \cite{vdp89})
\begin{subequations}
\begin{align}
        \bar{\Lambda}^-(x)&=\sum_{i=1}^{\infty}\int_{\mathbb R^2}
        \delta_{\tfrac{l'+a}{2}}(x-x')\bar{w}_i(x',l')\bar{\Xi}^-(x',l')
        \,d(x',l'),\\
        \bar{\Lambda}^+(x)&=\sum_{i=1}^{\infty}\int_{\mathbb R^2}
        \bar{\Xi}^+(x',l')\bar{w}_i(x',l')\delta_{\tfrac{l'+a}{2}}(x'-x)
        \,d(x',l'),
\end{align}
\end{subequations}
as well as
\begin{equation}
        \bar{\Xi}^-(x)=1+\int_{-\infty}^x\bar{\Lambda}^-(y)\,dy,\quad
        \bar{\Xi}^+(x)=1+\int_x^{+\infty}\bar{\Lambda}^+(y)\,dy.
\end{equation}
Then we find by \eqref{n.rod} and \eqref{n.pff.od} the reduction (cf.\
\cite{vdp89,bp96})
\begin{gather}\label{n.dr.od}
        \bar{\Xi}^{\pm}(x,l)=\bar{\Xi}^{\pm}(x\pm\tfrac{l+a}{2}),
\intertext{and via \eqref{n.bd} that}
        \dfrac{\bar{\Lambda}^-(x)}{\Xi}=\dfrac{\varrho^-(x)}{\bar{\Xi}^+(x)},
        \quad
        \dfrac{\bar{\Lambda}^+(x)}{\Xi}=\dfrac{\varrho^+(x)}{\bar{\Xi}^-(x)}
\intertext{with the densities at the boundaries of the blocks of $i$
interacting particles}\label{n.m.od}
        \varrho_i^{\pm}(x)=\int_{\mathbb R}\varrho_i(x\pm\tfrac{l+a}{2},l)\,dl
\intertext{and}\label{n.ms.od}
        \varrho^{\pm}(x)=\sum_{i=1}^{\infty}\varrho_i^{\pm}(x).
\end{gather}
Hence, on using the boundary conditions \eqref{n.bc.od}, we infer that
(cf.\ \cite{vdp89})
\begin{equation}\label{n.m.m.od}
        \dfrac{\bar{\Xi}^+(x)}{\Xi}\dfrac{d\bar{\Xi}^-(x)}{dx}=\varrho^-(x),
        \quad
        -\dfrac{\bar{\Xi}^-(x)}{\Xi}\dfrac{d\bar{\Xi}^+(x)}{dx}=\varrho^+(x).
\end{equation}  
If we substract these two equations, then we obtain
\begin{equation}
        \dfrac{1}{\Xi}\dfrac{d[\bar{\Xi}^-(x)\bar{\Xi}^+(x)]}{dx}=
        \varrho^-(x)-\varrho^+(x),
\end{equation}
which integrates upon using the boundary conditions \eqref{n.bc.od} via
\eqref{n.dr.od} to
\begin{equation}\label{n.o.od}
        \dfrac{\bar{\Xi}^-(x)\bar{\Xi}^+(x)}{\Xi}=1+
        \int_{-\infty}^x[\varrho^-(y)-\varrho^+(y)]\,dy
        \equiv\omega(x).
\end{equation}
Thus we deduce from \eqref{n.bc.od}, \eqref{n.dr.od}, \eqref{n.m.m.od},
and \eqref{n.o.od} that (cf.\ \cite{vdp89,bp96})
\begin{gather}\label{n.pff.ir.od}
        \ln\bar{\Xi}^-(x)=
        \int_{-\infty}^x\dfrac{\varrho^-(y)}{\omega(y)}\,dy,
        \quad
        \ln\dfrac{\bar{\Xi}^+(x)}{\Xi}=-\int_{-\infty}^x
        \dfrac{\varrho^+(y)}{\omega(y)}\,dy,
\intertext{and so we conclude by \eqref{n.bc.od} via \eqref{n.dr.od} that}
        \ln\Xi=\int_{\mathbb R}\dfrac{\varrho^-(y)}{\omega(y)}\,dy
        =\int_{\mathbb R}\dfrac{\varrho^+(y)}{\omega(y)}\,dy,
\intertext{or more symmetrically}\label{n.hrmf.od}
        \ln\Xi=\dfrac{1}{2}\int_{\mathbb R}
        \dfrac{\varrho^-(y)+\varrho^+(y)}{\omega(y)}\,dy,
\end{gather}
which generalizes the results (3.10) respective (4.30) of \cite{kr92} for
dimerizing hard rods respective sticky cores to systems with arbitrary $V$.

Next, let us analyze the one-dimensional marginals $\varrho_i^{\pm}$ of
$\varrho_i$. According to \eqref{n.m.od} and \eqref{n.bd}, we have by
\eqref{n.pff.od} and \eqref{n.bf.od}
\begin{subequations}
\begin{align}
        \varrho_i^-(x+\tfrac{a}{2})&=\dfrac{1}{\Xi}\bar{\Xi}^+(x,0)\cdot
        \int_{\mathbb R}z(fz)^{i-1}(x,x-l)\,\bar{\Xi}^-(x-\tfrac{l}{2},l)
        \,dl,\\
        \varrho_i^+(x-\tfrac{a}{2})&=\dfrac{1}{\Xi}
        \int_{\mathbb R}\bar{\Xi}^+(x+\tfrac{l}{2},l)\,z(fz)^{i-1}(x+l,x)\,dl
        \cdot\bar{\Xi}^-(x,0).
\end{align}
\end{subequations}
Hence, on using the reparameterizations
\begin{subequations}
\begin{align}
        \int_{\mathbb R}z(fz)^{i-1}(x,x-l)\,\bar{\Xi}^-(x-\tfrac{l}{2},l)\,dl&=
        \int_{\mathbb R}z(fz)^{i-1}(x,y)\,\bar{\Xi}^-(y,0)\,dy,\\
        \int_{\mathbb R}\bar{\Xi}^+(x+\tfrac{l}{2},l)\,z(fz)^{i-1}(x+l,x)\,dl&=
        \int_{\mathbb R}\bar{\Xi}^+(y,0)\,(zf)^{i-1}z(y,x)\,dy,
\end{align}
\end{subequations}
as well as \eqref{n.dr.od}, we infer the block densities equations
\begin{subequations}\label{n.bde.od}
\begin{align}
        \varrho_i^-(x)&=\dfrac{1}{\Xi}\bar{\Xi}^+(x)\cdot z(x-\tfrac{a}{2})
        \cdot\int_{\mathbb R}(fz)^{i-1}(x-\tfrac{a}{2},y)
        \,\bar{\Xi}^-(y-\tfrac{a}{2})\,dy,\\
        \varrho_i^+(x)&=\dfrac{1}{\Xi}
        \int_{\mathbb R}\bar{\Xi}^+(y+\tfrac{a}{2})
        \,(zf)^{i-1}(y,x+\tfrac{a}{2})\,dy\cdot z(x+\tfrac{a}{2})
        \cdot\bar{\Xi}^-(x).
\end{align}
\end{subequations}
In particular, we find for $i=1$ the profile equation (cf.\ \cite{bp96})
\begin{equation}\label{n.pe.od}
        \varrho_1(x)=\varrho_1^{\pm}(x\mp\tfrac{a}{2})=
        \dfrac{1}{\Xi}\bar{\Xi}^+(x+\tfrac{a}{2})z(x)
        \bar{\Xi}^-(x-\tfrac{a}{2})
	\equiv z(x)\psi(x+\tfrac{a}{2},x-\tfrac{a}{2}),
\end{equation}
where the two-dimensional weight function $\psi$ (cf.\ \cite{bp96,p97})
evaluates by \eqref{n.o.od} and \eqref{n.pff.ir.od} for $y\geq x$ to
(cf.\ \cite{bp96})
\begin{gather}\label{n.wf.od}
	\psi(y,x)=\omega(y)\exp\Big[
	-\int_x^y\dfrac{\varrho^-(z)}{\omega(z)}\,dz\Big]
	=\omega(x)\exp\Big[
	-\int_x^y\dfrac{\varrho^+(z)}{\omega(z)}\,dz\Big]
\intertext{or}
	\psi(y,x)=\sqrt{\omega(y)\omega(x)}\exp\Big[-\dfrac{1}{2}
	\int_x^y\dfrac{\varrho^-(z)+\varrho^+(z)}{\omega(z)}\,dz\Big].
\end{gather}
Thus, upon substituting \eqref{n.pe.od} into \eqref{n.bde.od}, and since
$\varrho_1$ is the density of the monomers, result \eqref{n.bde.od}
characterizes blocks as conformations of monomers correlated by the attractive
part of the interaction. This proposition, as the foregoing hard rod mixture
form density functional \eqref{n.hrmf.od}, indicates a general pattern of
transition from $D=1$ to $D>1$ (cf.\ Section \ref{c}).

Finally, to close the system of equations for the local particle density, we
derive a subsidiary condition for the marginals $\varrho_i^{\pm}$ in terms of
$\varrho$. Let us first rewrite \eqref{n.pd} in the form
\begin{align}\label{n.pd.sd.od}
        \varrho(x)=\dfrac{\delta\ln\Xi}{\delta\ln z(x)}
        &=\sum_{i=1}^{\infty}\int_{\mathbb R^2}
        \dfrac{\delta\ln\Xi}{\delta\ln\bar{w}_i(y,l)}
        \dfrac{\delta\ln\bar{w}_i(y,l)}{\delta\ln z(x)}\,d(y,l)\nonumber\\
        &=\int_{\mathbb R^2}
        \dfrac{\delta\ln\Xi}{\delta\ln[\sum_{i=1}^{\infty}\bar{w}_i(y,l)]}
        \dfrac{\delta\ln[\sum_{i=1}^{\infty}\bar{w}_i(y,l)]}{\delta\ln z(x)}
        \,d(y,l).
\end{align}
This yields
\begin{equation}
        \varrho(x)=\dfrac{1}{\Xi}\int_{\mathbb R^2}
        \bar{\Xi}^+(y,l)\,(\mathbb{I}-zf)^{-1}(y+\tfrac{l}{2},x)\,z(x)
        \,(\mathbb{I}-fz)^{-1}(x,y-\tfrac{l}{2})\,\bar{\Xi}^-(y,l)\,d(y,l)
\end{equation}
via \eqref{n.tgm.od.df}, \eqref{n.bd}, and \eqref{n.la.od}. Accordingly, we
have through a reparameterization of the integral and by \eqref{n.pd} as well
as \eqref{n.dr.od} that \cite{bp96}
\begin{subequations}\label{n.pff.i.sd.od}
\begin{align}
        \bar{\Xi}^-(x-\tfrac{a}{2})&=
        \int_{\mathbb R}(\mathbb{I}-fz)(x,y)\,\Xi^-(y)\,dy,\\
        \bar{\Xi}^+(x+\tfrac{a}{2})&=
        \int_{\mathbb R}\Xi^+(y)\,(\mathbb{I}-zf)(y,x)\,dy,
\end{align}
\end{subequations}
or that
\begin{subequations}\label{n.pff.sd.od}
\begin{align}
        \Xi^-(x)&=\int_{\mathbb R}(\mathbb{I}-fz)^{-1}(x,y)
        \,\bar{\Xi}^-(y-\tfrac{a}{2})\,dy,\\
        \Xi^+(x)&=\int_{\mathbb R}\bar{\Xi}^+(y+\tfrac{a}{2})
        \,(\mathbb{I}-zf)^{-1}(y,x)\,dy.
\end{align}
\end{subequations}
Hence, on combining \eqref{n.pff.sd.od} with the block densities equations
\eqref{n.bde.od}, we obtain \cite{bp96}
\begin{subequations}\label{n.m.m.sd.od}
\begin{align}
        \varrho^-(x+\tfrac{a}{2})&=\dfrac{1}{\Xi}
        \bar{\Xi}^+(x+\tfrac{a}{2})z(x)\Xi^-(x),\\
        \varrho^+(x-\tfrac{a}{2})&=\dfrac{1}{\Xi}
        \Xi^+(x)z(x)\bar{\Xi}^-(x-\tfrac{a}{2}),
\end{align}
\end{subequations}
and so we conclude via \eqref{n.pd}, \eqref{n.m.m.sd.od}, and \eqref{n.pe.od}
that (cf.\ \cite{bp96})
\begin{equation}\label{n.sc.od}
        \varrho(x)=\dfrac{\varrho^+(x-\tfrac{a}{2})\varrho^-(x+\tfrac{a}{2})}
        {\varrho_1(x)},
\end{equation}
which is the desired closure relation. On starting from \eqref{n.pd.sd.od}, the
very simple formal structure of \eqref{n.sc.od} is certainly due to the $D=1$
next neighbor nature of our problem. Therefore, the form \eqref{n.pd.sd.od}
has a higher dimensional analogue [as the result \eqref{n.bde.od} respective
\eqref{n.hrmf.od}], whereas \eqref{n.sc.od} has not. Proposition
\eqref{n.sc.od} completes the deduction of the nonuniform case.

In passing, we point out that the main results of our analysis are completely
characterized by the pressure-like quantities, termed effective pressures,
$p^{\pm}$, defined as \cite{p76,z91}
\begin{equation}\label{n.lp.od}
        \beta p^{\pm}(x)=\dfrac{\varrho^{\pm}(x)}{\omega(x)}
\end{equation}
with the $\varrho^{\pm}$ brought in via \eqref{n.ms.od} and $\omega$ by
\eqref{n.o.od}. We also introduce
\begin{equation}\label{n.ilp.od}
	\beta\pi^{\pm}(y,x)=\int_x^y\beta p^{\pm}(z)\,dz.
\end{equation}
Upon using definition \eqref{n.lp.od} and the corresponding results in the
form derived so far, one can show that our overcomplete format free energy $F$
for $D=1$ particle systems with next neighbor interactions is given by
\begin{gather}
        \beta F[U]=\min_{\varrho\in D_n}\bigg\{\int_{\mathbb R}
        \varrho(x)[\beta U(x)+\ln z(x)]\,dx-\ln\Xi\bigg\}
\intertext{with $D_n=\left\{t\in L^1(\mathbb{R}):\,t(x)\geq 0\wedge
\int_\mathbb{R}t(x)\,dx=n\right\}$, $n$ fixed, the profile equation}
        \beta[\mu-U(x)]=\ln\varrho_1(x)-\dfrac{1}{2}\ln[\psi^+(x)\psi^-(x)],
\intertext{where $\psi^{\pm}(x)=\omega(x\mp\tfrac{a}{2})
\exp[-\beta\pi^{\pm}(x+\tfrac{a}{2},x-\tfrac{a}{2})]$, 
$\psi^-\equiv\psi^+$ by \eqref{n.wf.od}, and the total Gibbs measure of hard
rod mixture form (cf.\ \cite{z91})}\label{n.hrmf.pr.od}
        \ln\Xi=\dfrac{1}{2}\int_{\mathbb R}[\beta p^-(y)+\beta p^+(y)]\,dy.
\end{gather}
One further finds that the block densities equations transcribe to
\begin{subequations}\label{n.bde.pr.od}
\begin{align}
        \varrho_i^-(x)&=\varrho_1^-(x)\cdot\int_\mathbb{R}
        \Big(f\dfrac{\varrho_1}{\psi^-}\Big)^{i-1}
        (x-\tfrac{a}{2},y+\tfrac{a}{2})\,\exp[-\beta\pi^-(x-a,y)]\,dy,\\
        \varrho_i^+(x)&=\int_\mathbb{R}\exp[-\beta\pi^+(y,x+a)]
        \,\Big(\dfrac{\varrho_1}{\psi^+}f\Big)^{i-1}
        (y-\tfrac{a}{2},x+\tfrac{a}{2})\,dy\cdot\varrho_1^+(x)
\end{align}
\end{subequations}
with the subsidiary condition \eqref{n.sc.od}. These results
extend---indirectly---the well-known (see, e.g., \cite{r69})
isothermal-isobaric technique from the uniform to the nonuniform case. So we
are led to conjecture that a direct nonuniform isothermal-isobaric approach
should be possible as well.

\subsection{Examples}

We apply the theory to the adhesive respective HGM system. This allows us to
make contact with the associated results for the uniform case, Section
\ref{uc}, and to provide detailed evidence for our conjecture on the
nonuniform isothermal-isobaric format.

To present the examples most adequately, we give first closed expressions for
the sums $\varrho^{\pm}$. Again, these simplifications are only possible for
the $D=1$ next neighbor case. Namely, if we start with \eqref{n.m.m.sd.od},
and use \eqref{n.pff.i.sd.od} as well as \eqref{n.pe.od}, we end up with
\begin{subequations}\label{n.ms.ce.od}
\begin{align}
        \dfrac{\varrho(x-\tfrac{a}{2})}{\varrho^-(x)}&=
        1+\int_{\mathbb R}f(y+a-x)\beta p^+(y)\exp[-\beta\pi^+(y,x)]\,dy,\\
        \dfrac{\varrho(x+\tfrac{a}{2})}{\varrho^+(x)}&=
        1+\int_{\mathbb R}f(x+a-y)\beta p^-(y)\exp[-\beta\pi^-(x,y)]\,dy
\end{align}
\end{subequations}
via the subsidiary condition \eqref{n.sc.od}, \eqref{n.m.m.od}, and through
definition \eqref{n.lp.od}. This result is equivalent to the one-component
case of formula (3.12) of \cite{bp96}.

\subsubsection{Adhesive interaction}

Due to the contact nature of \eqref{ap.ai}, we find by \eqref{n.ms.ce.od}
\begin{subequations}\label{n.ms.ce.ai.od}
\begin{align}
        \dfrac{\varrho(x-\tfrac{a}{2})}{\varrho^-(x)}&=
        1+\lambda\beta p^+(x),\\
        \dfrac{\varrho(x+\tfrac{a}{2})}{\varrho^+(x)}&=
        1+\lambda\beta p^-(x),
\end{align}
\end{subequations}
which are the generalizations of \eqref{u.od.ai.mc} to the nonuniform case.
Consequently, we have that
\begin{equation}\label{n.o.ai.od}
        \omega(x)=1-\int_{x-\tfrac{a}{2}}^{x+\tfrac{a}{2}}\varrho(y)\,dy
        \equiv 1-\tau(x),
\end{equation}
and by \eqref{n.sc.od} also the case $i=1$ of \eqref{u.od.ai.c} at the
nonuniform level,
\begin{equation}\label{n.sc.ai.od}
        \varrho_1(x)=\dfrac{\varrho(x)}{[1+\lambda\beta p^-(x-\tfrac{a}{2})]
        [1+\lambda\beta p^+(x+\tfrac{a}{2})]}.
\end{equation}

Moreover, rewriting \eqref{n.ms.ce.ai.od} via \eqref{n.lp.od} and
\eqref{n.o.ai.od} as
\begin{subequations}\label{n.ms.ce.pe.ai.od}
\begin{align}
        \dfrac{\varrho(x-\tfrac{a}{2})}{1-\tau(x)}&=
        \beta p^-(x)[1+\lambda\beta p^+(x)],\\
        \dfrac{\varrho(x+\tfrac{a}{2})}{1-\tau(x)}&=
        \beta p^+(x)[1+\lambda\beta p^-(x)]
\end{align}
\end{subequations}
gives
\begin{equation}\label{n.lp.ai.od}
        \beta p^{\pm}(x)=\dfrac{1}{2\lambda}
        \Bigg\{-1\pm\dfrac{\tau'(x)}{1-\tau(x)}+
        \sqrt{1+\dfrac{4\lambda\sigma(x)}{1-\tau(x)}+
        \bigg[\dfrac{\lambda\tau'(x)}{1-\tau(x)}\bigg]^2}\Bigg\}
\end{equation}
with $\sigma(x)=\tfrac{1}{2}[\varrho(x+\tfrac{a}{2})+\varrho(x-\tfrac{a}{2})]$.
Hence, on substituting \eqref{n.lp.ai.od} into \eqref{n.hrmf.pr.od}, we
arrive at formula (5.17) respective (6.16) of \cite{p82}. In the uniform
limit, we recover \eqref{u.eos.ai.i} respective \eqref{u.eos.ai}.

Finally, if we specialize \eqref{n.bde.pr.od} to \eqref{ap.ai}, then we obtain
via \eqref{n.o.ai.od} that
\begin{subequations}
\begin{align}
        \varrho_i^-(x)&=\varrho_1^-(x)\,\lambda^{i-1}
        \prod_{j=1}^{i-1}\dfrac{\varrho_1^-(x-ja)}{1-\tau(x-ja)},\\
        \varrho_i^+(x)&=\varrho_1^+(x)\,\lambda^{i-1}
        \prod_{j=1}^{i-1}\dfrac{\varrho_1^+(x+ja)}{1-\tau(x+ja)},
\end{align}
\end{subequations}
which correspond immediately to \eqref{u.od.ai.c} for $i>1$ through
\eqref{n.sc.ai.od} and \eqref{n.ms.ce.pe.ai.od}.

\subsubsection{HGM system}

For an interaction characterized by \eqref{ap.hgm}, the block densities
equations \eqref{n.bde.pr.od} are no longer analytically tractable. Thus,
upon substituting \eqref{ap.hgm} into \eqref{n.ms.ce.od}, we are left with
\begin{subequations}\label{n.ms.ce.hgm.od}
\begin{align}
        \dfrac{\varrho(x-\tfrac{a}{2})}{\varrho^-(x)}&=
        1+\lambda\{1-\exp[-\beta\pi^+(x+d,x)]\},\\
        \dfrac{\varrho(x+\tfrac{a}{2})}{\varrho^+(x)}&=
        1+\lambda\{1-\exp[-\beta\pi^-(x,x-d)]\},
\end{align}
\end{subequations}
and so we have by \eqref{n.sc.od} that
\begin{align}
        \dfrac{\varrho(x)}{\varrho_1(x)}&=\big(1+\lambda
        \{1-\exp[-\beta\pi^-(x-\tfrac{a}{2},x-\tfrac{a}{2}-d)]\}
	\big)\nonumber\\
        &\times\big(1+\lambda
        \{1-\exp[-\beta\pi^+(x+\tfrac{a}{2}+d,x+\tfrac{a}{2})]\}\big).
\end{align}
These results generalize \eqref{u.od.hgm.mc} respective \eqref{u.od.hgm.c} for
$i=1$ to the nonuniform case.

Furthermore, on substituting \eqref{n.ms.ce.od} into \eqref{n.o.od}, we find
through \eqref{ap.hgm}, \eqref{n.wf.od}, as well as definitions
\eqref{n.lp.od} respective \eqref{n.ilp.od}
\begin{equation}
	\omega(x)=1-\tau(x)-\lambda\int_x^{x+d}\varrho^+(y)
	\bigg\{\int_{y-d}^x\beta p^-(z)\exp[-\beta\pi^-(y,z)]\,dz\bigg\}\,dy,
\end{equation}
yielding via definition \eqref{n.ilp.od}, \eqref{n.ms.ce.hgm.od},
\eqref{n.wf.od}, and definition \eqref{n.lp.od}
\begin{align}
	\omega(x)&=1-\tau(x)+\int_x^{x+d}\cfrac{\varrho(y+\tfrac{a}{2})}
        {\exp[\beta\pi^-(y,y-d)]\bigg(1+\cfrac{1}{\lambda}\bigg)-1}\,dy
	\nonumber\\
	&-\lambda\int_x^{x+d}\beta p^+(y)\omega(x)\exp[-\beta\pi^+(y,x)]\,dy,
\end{align}
and so we conclude on using again proposition \eqref{n.ms.ce.hgm.od} that
the effective pressure $p^-$ is determined by [cf.\ for the uniform case
\eqref{u.eos.hgm.i}]
\begin{subequations}
\begin{align}
        1-\tau(x)&=\dfrac{\varrho(x-\tfrac{a}{2})}{\beta p^-(x)}-
        \int_x^{x+d}\cfrac{\varrho(y+\tfrac{a}{2})}
        {\exp[\beta\pi^-(y,y-d)]\bigg(1+\cfrac{1}{\lambda}\bigg)-1}\,dy,
\intertext{and likewise $p^+$ by}
        1-\tau(x)&=\dfrac{\varrho(x+\tfrac{a}{2})}{\beta p^+(x)}-
        \int_{x-d}^x\cfrac{\varrho(y-\tfrac{a}{2})}
        {\exp[\beta\pi^+(y+d,y)]\bigg(1+\cfrac{1}{\lambda}\bigg)-1}\,dy,
\end{align}
\end{subequations}
which is equivalent to formula (6.15) of \cite{bp96}. This completes our
discussion of the nonuniform HGM system within the $p^{\pm}$ representation.

In summary, the examples demonstrate in detail that a suggestive formal
correspondence between the usual uniform isothermal-isobaric method and the
nonuniform effective pressures format exists. This supports the idea of an
isothermal-isobaric representation for arbitrary nonuniform $D=1$ particle
systems.

\section{Conclusion}\label{c}

We presented a realization of the overcompleteness strategy for $D=1$ systems
with next neighbor interactions, where the supplementary variables are the
local block densities $\varrho_i$ of $i$ associated particles. The
corresponding free energy density functional is of the simple additive hard
rod mixture type. The isomorphism between our overcomplete format and the
standard description was established through a trivial identity of functional
analysis. The first results on this class of density functionals have been
derived by Kierlik and Rosinberg \cite{kr92} for the special case of
dimerizing hard rods. So we lifted their approach to particle systems with
arbitrary next neighbor interactions. Moreover, we showed that all our main
results are completely characterized by the effective pressures $p^{\pm}$.
This proposition in conjunction with the detailed results for the adhesive
interaction and the HGM system support the idea of a nonuniform
isothermal-isobaric representation (analoguous to the well-known uniform
isothermal-isobaric method) for arbitrary nonuniform $D=1$ systems.

To introduce our concept of overcomplete description most concisely, we
started with the uniform case. We treated first the two examples of adhesive
systems and HGM systems. These results formed also the basis for our conjecture
on a nonuniform isothermal-isobaric representation. We deduced our overcomplete
description for arbitrary next neighbor interactions by means of an application
of large deviation techniques, accompanied by the Hardy-Ramanujan formula on
the number of partitions of a positive integer.

What we can learn from the $D=1$ case is the logic of the problem for $D>1$
systems with finite range interactions. In abstracting from the block densities
equations \eqref{n.bde.pr.od}, (i) we decompose a given $D>1$ system into
blocks of $i$ monomers correlated by the attractive part of the interaction.
(ii) Hence, on interpreting the blocks as components of a mixture, where the
interaction between the components is given by construction by a hard core
potential, we arrive at the extension of the hard rod mixture form density
functional \eqref{n.hrmf.pr.od}. (iii) Finally, we establish the isomorphism
between our overcomplete description and the standard formalism by an
application of an identity of functional analysis as in \eqref{n.pd.sd.od}. A
familiar strategy to organize such a first-principles approach is to reduce
the geometry of the phase space by going over to the corresponding discrete
systems (see, e.g., \cite{r69}). In particular, the prototypical simply
connected infinite lattices, along the Cayley tree (for an overview see, e.g.,
\cite{p96}), can be used to demonstrate the effectivity of our method. Progress
in this direction will be the subject of a future communication.

Complementary to the reduction of the geometry of the phase space, another
possibility when trying to apply mathematical techniques is to restrict the
class of interactions. For the prominent example of systems with strong, short
range interactions, free energy models can be constructed along these lines.
We begin with the physical observation, that in such, e.g., $D=3$ systems the
blocks take on predominantly elementary sphere-like or cube-like
conformations. Thus, the available analytic expressions for the hard sphere
respective hard cube mixture free energy functional can be used to model the
hard core interaction between the blocks [cf.\ (ii)]. Modelling of the internal
block free energy [cf.\ (i)] and the incorporation of the subsidiary condition
[cf.\ (iii)] requires a more incisive approach, but is still amenable due to
the strong, short range character of the interaction. Also the implementation
of this program will be reported in the future.

\providecommand{\bysame}{\leavevmode\hbox to3em{\hrulefill}\thinspace}

\end{document}